\documentclass[conference]{IEEEtran}
\IEEEoverridecommandlockouts

\usepackage{cite}
\usepackage{amsmath,amssymb,amsfonts}
\usepackage{algorithmic}
\usepackage{graphicx}
\usepackage{textcomp}
\usepackage{xcolor}

\usepackage{booktabs} % For formal tables
\usepackage{listings}
\usepackage{ifthen}
\usepackage{color}
\usepackage{ifthen}
\usepackage{soul}
\usepackage{balance}

\definecolor{dkred}{rgb}{.6,0,0}
\definecolor{dkgreen}{rgb}{0,.5,0}
\definecolor{dkblue}{rgb}{0,0,.6}
\definecolor{dkyellow}{cmyk}{0,0,.8,.3}
\definecolor{lightgray}{rgb}{.95,.95,.95}
\definecolor{darkgray}{rgb}{.3,.3,.3}
\definecolor{darkblue}{rgb}{0,0,.20}
\definecolor{purple}{rgb}{0.65, 0.12, 0.82}

\newcommand{\code}[1]{\mbox{\lstinline@#1@}}
\newcommand\stitle[1]{\noindent\textbf{#1.}}

\lstdefinestyle{PHP} {
	language=php,
	basicstyle = \footnotesize\ttfamily,
}

\lstdefinelanguage{code}
  { 
    keywords={},
    otherkeywords={var_dump},
	basicstyle=\fontsize{8pt}{8}\selectfont\ttfamily,
    keywordstyle=\bfseries\color{blue},
    sensitive=false,
    commentstyle=\color{purple!40!black},
    showspaces=false,
    numbers=left,
    numbersep=3pt,  
    tabsize=1,
    literate= {~}{\texttt{\phantom{m}}}1 {`}{$\texttt{\%}$}1 {±}{$\texttt{\$}$}1, %{~}{$\sim$}{1} 
    showstringspaces=false,
    emph={3},
    showtabs=true,
    morecomment=[l]{//},
    morecomment=[s]{/*}{*/},
    morestring=[b],
    breaklines=true,
    breakindent=12pt
  }
\let\mmp\lstinline

\newcommand\tk[1]{{\small \textsf{#1}}}
\newcommand\cmd[1]{{\small \texttt{#1}}}
\newcommand\DT{{DeepTective}}

\newcommand{\gitfi}{GIT}
\newcommand{\gitfu}{GIT$^f$}
\newcommand\combofu{ALL$^f$}
\newcommand\combofi{ALL}

\begin{document}
\title{A Hybrid Graph Neural Network Approach for Detecting PHP Vulnerabilities
\thanks{A poster version of this paper appeared as~\cite{sec-at-sac}.}}

\author{\IEEEauthorblockN{Rishi Rabheru}
\IEEEauthorblockA{\textit{Department of Computing} \\
\textit{Imperial College London}\\
rishi.rabheru16@imperial.ac.uk}
\and
\IEEEauthorblockN{Hazim Hanif}
\IEEEauthorblockA{\textit{Department of Computing} \\
\textit{Imperial College London;}\\
\textit{Faculty of Computer Science}\\ 
\textit{and Information Technology}\\
\textit{University of Malaya, Malaysia}\\
m.md-hanif19@imperial.ac.uk}
\and
\IEEEauthorblockN{Sergio Maffeis}
\IEEEauthorblockA{\textit{Department of Computing} \\
\textit{Imperial College London}\\
sergio.maffeis@imperial.ac.uk}
}

\maketitle

\begin{abstract}
    This paper presents DeepTective, a deep learning approach to detect vulnerabilities in PHP source code. Our approach implements a novel hybrid technique that combines Gated Recurrent Units and Graph Convolutional Networks to detect SQLi, XSS and OSCI vulnerabilities leveraging both syntactic and semantic information. We evaluate DeepTective and compare it to the state of the art on an established synthetic dataset and on a novel real-world dataset collected from GitHub. 
    Experimental results show that DeepTective achieves near perfect classification on the synthetic dataset, and an F1 score of 88.12\% on the realistic dataset, outperforming related approaches. 
    We validate DeepTective in the wild by discovering 4 novel vulnerabilities in established WordPress plugins.
\end{abstract}

\section{Introduction}

PHP remains the most common server-side language on the web, especially among the long tail of medium and small size websites.
Due to the amount of economic activity taking place online, PHP web applications remain a tempting target for malicious actors looking to exploit security vulnerabilities for financial gain or in pursue of other illicit ends.
In order to preempt the compromise of PHP web applications there has been a steady and growing trend by developers, security firms and white hat hackers to find, fix and disclose PHP vulnerabilities~\cite{MITRE2020BrowseDate}.

The research community has also devoted a significant amount of effort to the automated discovery of PHP vulnerabilities. Besides established approaches based on static, flow, and taint analysis~\cite{Pixy2006,Dahse2010rips,DBLP:conf/ndss/DahseH14} data mining has proven to be another effective approach~\cite{Son2011SAFERPHP,Medeiros2014WAP,medeiros_equipping_2016,Nunes2015PhpSafe,huang_uchecker_2019}.
These solutions are very efficient in analysing large quantities of code, but tend to suffer from limited detection performance, in terms of false positives or false negatives. 
Following recent advances in deep learning and natural language processing, security researchers started to develop deep learning based approaches to detect software vulnerabilities in C and C++ programs~\cite{Russell2018Draper,li2018vuldeepecker}. 
Only very recently we have seen the first applications of deep learning to PHP vulnerability discovery~\cite{Fang2019Tap,Fidalgo2020NewDekant,Guo2020Vulhunter}.
Both these approaches apply Long-Short Term Memory (LSTM) neural networks to capture non-local dependencies over various transformations of the source code.
LSTM is good at finding patterns in sequential data but is not equally well suited to learn from tree- or graph-structured data, which is a more natural representation of program semantics.

In this paper we present \DT, a deep-learning based vulnerability detection approach, which aims to combine both syntactic and semantic properties of source code.

The first key decision in a machine learning based vulnerability detection pipeline for source code is what code units constitute the samples.
Many options are possible, ranging from lines of code to entire code bases.
Depending on the technique employed, the characteristics of vulnerabilities targeted and the detection objectives, a number of options can be appropriate.
In our case we selected both function- and file-level granularities for which we collected novel datasets and explored the trade-offs between the two. 
Intuitively, function-level classification helps to better locate vulnerable code and reduces the amount of unrelated source code to consider during the learning phase, which improves efficiency and may increase precision.
On the other hand, some vulnerabilities are due to the interaction of non-local snippets of code, and may require a whole file, or even more, for identification. Moreover, file-level samples are easier to label in an automated fashion, and therefore constitute better learning datasets.

In order to learn syntactic and structural properties from source code, \DT\ transforms it into a sequence of tokens to be analysed by a Gated Recurrent Unit (GRU), a neural network related to the LSTM and able to embed sequential information, in this case about the code structure.
Novel to our approach for PHP, we attempt to learn semantic properties of the source code by analysing the CFG with a Graph Convolutional Network (GCN), a recent neural network architecture designed to handle graph-like data structures which during training can embed semantic and contextual information of the source code into the classification model. 
For our best model, this hybrid architecture achieves a 99.92\% F1 score on synthetic data from SARD~\cite{sardphp} and a 88.12\% F1 score on real-world data from GitHub (our novel dataset).

We investigate the impact of different dataset distributions for detecting multiple vulnerabilities, and the challenges in creating such datasets. 
The key dimensions to take into account are the nature of the samples (synthetic versus realistic), the accuracy of the labels, the balance of the classes and the overarching difficulty in generating high-quality datasets.
Existing work on PHP emphasised the use of clean and synthetic datasets, and in particular SARD. 
We found that even a model achieving 100\% F1-score when trained and tested on different portions of the same synthetic dataset can have dismal performance when tested on realistic data.
We systematically compare the performance of \DT\ and a number of existing PHP vulnerability detection tools on SARD and on our real-world dataset. 
\DT\ outperforms the other tools on both datasets, but the gap becomes extremely large on the real-world one, even for pre-trained models.

Finally, we tested \DT\ in the wild, evaluating its execution performance and its ability to generalise to a number of real-world PHP applications not present in the training dataset. We validated the practical usefulness of \DT\ by discovering 4 novel SQL injection and Cross-site scripting vulnerabilities in deployed plugins for WordPress. 

In summary, our main contributions are: 

\begin{itemize}
    \item The first investigation of the use of GCN and GRU to detect vulnerabilities in PHP source code, embedding both syntactic, structural and semantic information in the machine learning model.
    
    \item An analysis of the impact of dataset definition on model performance for vulnerability discovery, and the collection of new function- and file-level labelled PHP datasets.
    
    \item An extended evaluation of \DT, our GNN, by comparing it with selected existing tools for PHP vulnerability detection and by using it in the wild, where we discovered 4 novel vulnerabilities in established WordPress plugins.

\end{itemize}

\section{Background}

In this Section, we review the three common PHP vulnerabilities targeted by our detector, and survey automated vulnerability detection approaches for PHP source code.

\subsection{PHP Vulnerabilities}

A software vulnerability is a mistake in software made by the software developer that can be used by a malicious actor to gain access to a system or network \cite{MITRE2019FAQ}. The 2020 CWE Top 25 Most Dangerous Software Weakness \cite{MITRE2020Top} list different types of vulnerabilities that are frequently reported in the current web application ecosystem. Since we target PHP source code, we focus on three types of vulnerabilities particularly common in PHP web applications: SQL injection (SQLi), Cross-site scripting (XSS) and OS Command injection (OSCI). 
Table~\ref{tab:code_snippet} shows vulnerable snippets of code from our \gitfi\ dataset (described in Section~\ref{sec:dataset}) for SQLi, XSS and OSCI. These are real-world security vulnerabilities found on open source PHP applications hosted on GitHub. We highlighted the vulnerable parts of each code snippet, and we explain them below.

SQLi, also known as CWE-89, is a vulnerability that occurs when an attacker manages to alter a SQL query before it is passed to a database~\cite{Clarke2009sql}. 
It allows attackers to read or write data from the database without authorization, or to launch Denial of Service attacks. 
For example, in the first code snippet of Table \ref{tab:code_snippet}, the attacker is able to perform a SQLi by 
providing a carefully formatted string in place of their email in the variable \mmp$±email$, which is used as part of a dynamically generated SQL query in the highlighted line of code. 
The string provided by the attacker may be of the form ``\mmp$'' UNION [malicious SQL query]$'', where the attacker can use a succession of malicious SQL queries to learn about the structure of the database, exfiltrate and modify data.

XSS, or CWE-79, is a vulnerability where the web application fails to sanitize user input before displaying it on a web page \cite{MITRE2019CWE-79Scripting}. This vulnerability occurs due to a nonexistent or insufficient implementation of input and output sanitization, which allows attackers to inject arbitrary JavaScript into an HTML file \cite{Johns2008Xss}. A successful XSS can lead to a number of malicious activities including extracting confidential information, such as session cookies, or sending malicious requests to other web applications. Table \ref{tab:code_snippet} shows an example of an XSS vulnerability found in a real-world PHP project. The highlighted part indicates the vulnerable part of the code. Based on this example, the input from \mmp$POST$ is not sanitised before being printed on the same line. Therefore, an attacker can trick a victim into inserting an attacker-controlled \mmp$<script>$ tag in the page, implementing arbitrary client-side malicious behaviour. 

OSCI, or CWE-78 is a web application vulnerability that allows attackers to execute malicious operating systems commands on the targeted server that runs the vulnerable web application \cite{LIU2012Improving}. The severity of this type of attack depends on the privilege level that the attackers gains through the injection attack. The highlighted code in Table \ref{tab:code_snippet} shows an example of a real-world OSCI vulnerability where the filename from \mmp$FILES['file']['name']$ is concatenated with a string to form a full file path. It is intended to be used to move files through the execution of shell command \mmp$mv$. However, due to the lack of sanitisation on the input for the file name, an attacker can exploit this vulnerability by setting a malicious filename like ``\mmp$any_name.txt; [any shell command here]$" to run arbitrary commands on the server.

\begin{table*}[t]
    \caption{Vulnerable code snippets from GitHub projects.}
    \label{tab:code_snippet}
    \centering
    \begin{tabular}{p{5.6cm}p{5.6cm}p{5.6cm}} 
        \toprule
        \textbf{SQL Injection (SQLi)} & \textbf{Cross-site scripting (XSS)} & \textbf{OS Command Injection (OSCI)} \\
        \midrule
\begin{lstlisting}[language=PHP,escapechar=@,basicstyle=\fontsize{6pt}{6pt}\selectfont\ttfamily, belowskip=-1.2 \baselineskip, aboveskip=-0.3 \baselineskip]
$user_name = ($firstname AND $lastname) ? $firstname.' '.$lastname : '';
$user_email = ($email) ? $email : $this->getRandomString();
$user_color = ($color) ? $color : $this->random_color();

@\hl{\$query = 'SELECT id FROM '.\$this->table\_prefix.'users WHERE `email` = \''.\$user\_email.'\' LIMIT 1;';}@
$usercheck = $this->db->query($query);

if ( isset($usercheck[0]->id) )
{
	$user_id = $usercheck[0]->id;
	$user_email = $usercheck[1]->email;
}
\end{lstlisting} & \begin{lstlisting}[language=PHP,escapechar=@,basicstyle=\fontsize{7pt}{3pt}\selectfont\ttfamily, belowskip=-1.2 \baselineskip, aboveskip=-0.3 \baselineskip]
@\hl{print\_r(\$\_POST);}@
if($_POST) {
		if(isset($_POST['webdav_url'])) {
			OC_CONFIG::setValue('user_webdavauth_url', strip_tags($_POST['webdav_url']));
		}
}

$tmpl = new OC_Template( 'user_webdavauth', 'settings');
$tmpl->assign( 'webdav_url', OC_Config::getValue( "user_webdavauth_url" ));

return $tmpl->fetchPage();
}\end{lstlisting} & \begin{lstlisting}[language=PHP,escapechar=@, basicstyle=\fontsize{7pt}{3pt}\selectfont\ttfamily, belowskip=-1.2 \baselineskip, aboveskip=-0.3 \baselineskip]
$zip = "/tmp/" . @\hl{\$\_FILES['file']['name'];}@
@\hl{\$command = "mv " .\$\_FILES['file']['tmp\_name']." \$zip";}@
@\hl{exec(\$command,\$output=array(),\$res);}@
if ($res) {
    $this->errors[] = lang::translate('gallery_error_zip_mv');
    return false;
}

@\hl{\$command = "chmod 777 " . \$zip;}@
@\hl{exec(\$command,\$output=array(),\$res);}@
if ($res) {
    $this->errors[] = lang::translate('gallery_error_zip_chmod');
}
\end{lstlisting}\\
    \bottomrule
    \end{tabular}
    %\vspace{-4mm}
\end{table*}

\subsection{Detecting Vulnerabilities in PHP}

Researchers and practitioners, over the years, have developed many tools to detect vulnerabilities in PHP applications. 

\subsubsection{Traditional Approaches}
Traditional approaches focus on the use of static, semantic and taint analysis to locate vulnerabilities. 
Pixy~\cite{Pixy2006} implements flow-sensitive and context-sensitive data flow analysis to detect vulnerable components in a PHP web applications, mainly targeting XSS. That approach can be extended to the detection of other taint-style vulnerabilities such as SQLi and OSCI.

RIPS~\cite{Dahse2010rips,DBLP:conf/ndss/DahseH14} combines taint and static analysis to locate vulnerable program points in a PHP application. However, RIPS and Pixy are unable to analyze flaws that require the analysis of multiple files, or that depend on object-oriented features of PHP, limiting their effectiveness on current web applications. 
phpSAFE \cite{Nunes2015PhpSafe} performs a lexical and semantic analysis of code at the Abstract Syntax Tree (AST) level, before executing an inter-procedural analysis to follow the flow of tainted variables starting from the \mmp$main$ function. The authors report to be able to detect SQLi and XSS vulnerabilities with a lower false positive rate than RIPS and Pixy. 
Differently from previous approaches, SAFERPHP \cite{Son2011SAFERPHP} focuses on the detection of Denial of Service (DoS) and missing authorisation checks. Besides implementing taint analysis in the process, SAFERPHP also performs inter-procedural and semantic analysis by analysing the control dependencies via the control flow graph (CFG). The analysis allows the tool to identify and verify the consistency of possible security checks in all calling contexts.

\subsubsection{Data Mining Approaches}

More recent approaches aim to detect PHP web application vulnerabilities using data mining techniques.
WAP~\cite{Medeiros2014WAP,medeiros_equipping_2016} implements taint analysis along with a number of machine learning models to predict vulnerable PHP samples. Logistic Regression obtains the best performance, and is able to detect 8 classes of vulnerabilities, including SQLi, XSS, and OSCI. 
In follow up work, DEKANT~\cite{Medeiros2016Dekant} adopts Natural Language Processing (NLP) techniques to detect vulnerabilities. In particular, it uses a Hidden Markov Model (HMM)~\cite{Rabiner1989HMM} to characterise vulnerabilities based on a set of source code slices. These code slices are marked as tainted or non-tainted and then passed on for further analysis. Like WAP, DEKANT handles a number of different vulnerability classes. In a comparison against other tools, DEKANT achieved 96\% accuracy as compared to 90\% for WAP and 18\% for Pixy. 
%The rate of false positives and negatives of DEKANT is 12\%, and 0\% respectively, which are also lower than both of the previous approaches. Moving on to another tool,
%
WIRECAML~\cite{Kronjee2018Wirecaml} combines data-flow analysis and machine learning to detect SQLi and XSS vulnerabilities in PHP source code. The combination of reaching definition, taint and reaching constant analysis allows the tool to extract meaningful data flow features from the CFG, and optimise the learning processing of the machine learning model. The best results are obtained by a Decision Tree classifier with a precision-recall curve score of 88\% for SQLi and 82\% for XSS. In comparison with previous static analysis approaches like Pixy, RIPS and WAP, WIRECAML achieved better detection performance in terms of precision, recall and F1-scores on all cases except non-vulnerable XSS samples, where RIPS scored best. WIRECAML was used to detect a SQLi vulnerability in a photo gallery web application called Piwigo, allowing an attacker to inject arbitrary queries via a \mmp$POST$ parameter.

\subsubsection{Deep Learning Approaches}

More recently, deep learning is being applied to vulnerability detection for PHP source code.
TAP \cite{Fang2019Tap}  proposes a static analysis approach of detecting PHP vulnerabilities based on code tokens and deep learning techniques. The tool extracts code tokens from PHP codes using a custom tokenizer, and performs data flow analysis to find relevant lines of code that contain function calls. TAP uses Word2Vec to generate numerical vectors from the code tokens, and implements a sequence-based deep learning technique called Long Short-term Memory (LSTM) to train the detection model. TAP handles several classes of vulnerabilities, and in a comparison with WIRECAML and RIPS achieved the best results for accuracy, F1-score and area under the curve (AUC) on both safe and vulnerable samples.

Vulhunter \cite{Guo2020Vulhunter} proposes a different approach leveraging bytecode features to represent vulnerabilities. Vulhunter generates CFGs, data-flow graphs (DFGs) and analyses them to generate potentially suspicious code slices. The code slices are transformed into bytecode slices. Like TAP, Vulhunter uses Word2Vec to generate vectors from the bytecode slices. Vectors and tokens are passed to a Bi-directional Long Short-term Memory (Bi-LSTM). The evaluation results show that Vulhunter is capable of detecting SQLi and XSS vulnerabilities with higher recall and F1-scores than RIPS. Vulhunter was used to discover two XSS vulnerabilities and one SQLi vulnerability in SEACMS and CMS Made Simple. 

Also \cite{Fidalgo2020NewDekant} leverages PHP bytecode to locate vulnerabilities. Code slices are translated to bytecode using the Vulcan Logic Dumper (VLD), which intercepts Zend bytecode before it executed. Bytecode tokens are mapped to integers values understandable by a neural network using a vocabulary-based translation. The authors train a 2-layer LSTM model and achieved 95.35\% accuracy, 96.51\% precision and 96.14\% recall using RMSProp as the optimisation function during training. However, they only focus on detecting SQLi vulnerabilities and do not compare their performance with previous approaches.

To the best of our knowledge, we are the first to use graph neural networks for detecting vulnerabilities in PHP source codes, and to investigate the effect of synthetic versus realistic datasets on model performance.

\section{\DT}\label{sec:dt}

\begin{figure*}[!t]
 \centering
 \includegraphics[scale=0.12]{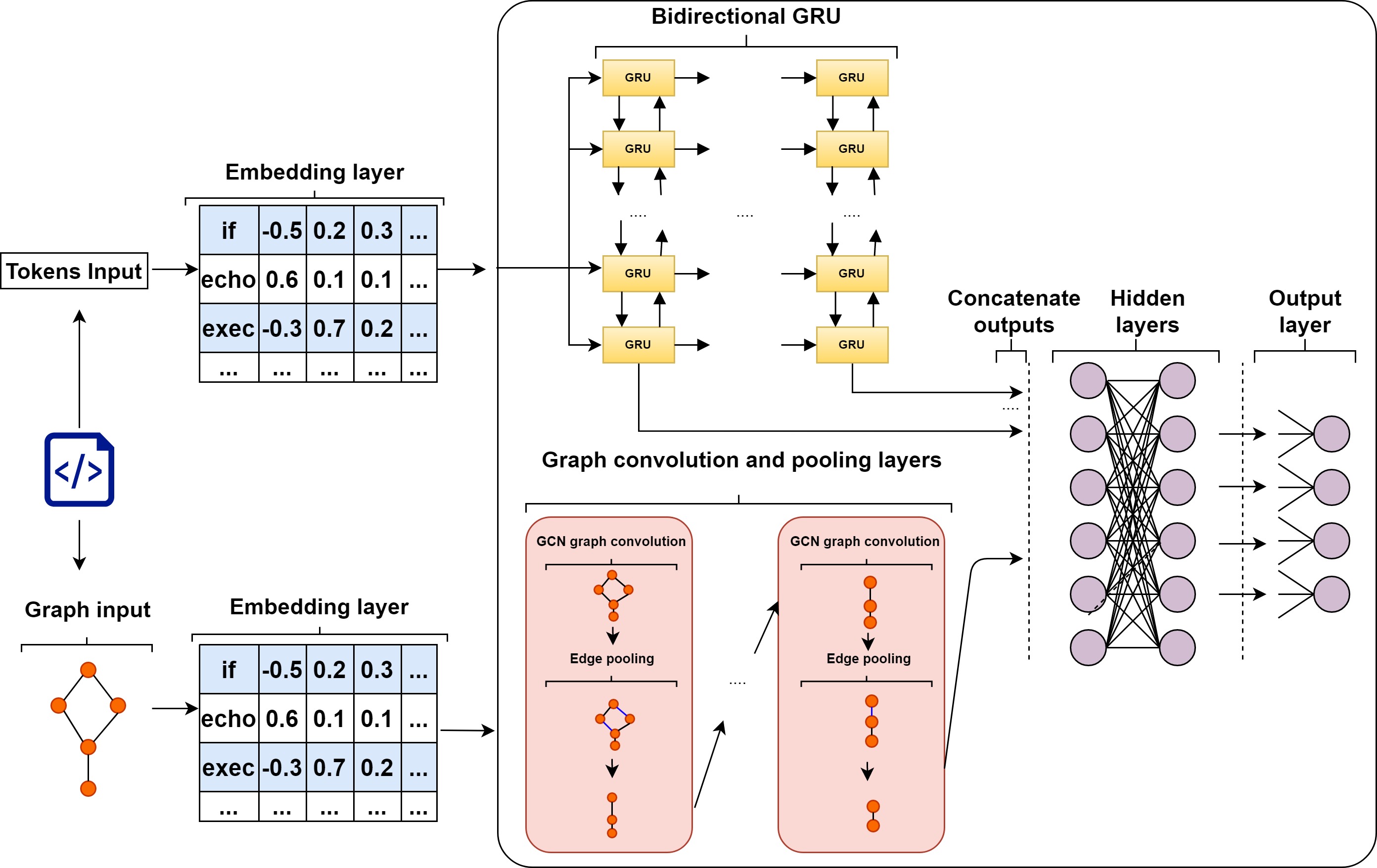} % was 0.09 in single column
    %\captionsetup{font=scriptsize}
    \caption{\DT\ Architecture}\label{fig:arch}
\end{figure*}

In this Section, we introduce \DT, our novel PHP vulnerability detection model. 
\DT\ detects SQLi, XSS and OSCI vulnerabilities within source code, at function- and file-level granularity. 
It is divided into two key components: a Gated Recurrent Unit (GRU) which operates on the linear sequence of source code tokens, and a Graph Convolutional Network (GCN) which operates on the Control Flow Graph (CFG) of the source code.
Each component provides a different mechanism for the model to detect multiple types of vulnerabilities effectively. 
We combine the GRU and GCN in a novel hybrid architecture able to leverage the strengths of both techniques. 

\subsection{Preprocessing}

Our data samples are fragments of PHP code, either a function body or a whole file. 
As a first step, we raise the level of abstraction of the code to a format that will conceptually help the learning process. 
We extract the linear sequence of parsed tokens in order to capture syntactic dependencies,
and we extract the set of intraprocedural CFGs to capture semantic dependencies.
We also transform the sequence of tokens and the CFG in a suitable format for consumption by a neural network, that is multidimensional vectors of real numbers.

\subsubsection{Sequence of Tokens}

We parse a sample using \cmd{phply}~\cite{phply}, a PHP parsing library built on top of \cmd{ply}~\cite{ply}, an implementation of the \cmd{yacc} and \cmd{lex} parsing tools for Python. 
From parsing, we obtain an ordered sequence of tokens. 
We remove tokens for comments, tabs, spaces, and PHP open and close tags from the sequence, as the presence or absence of a vulnerability is not affected by these.  

In order to focus the learning on a manageable set of interesting tokens, we conflate the long tail of user-defined functions, variables, and constant values into abstract tokens, and retain the concrete token only for the first $k$ instances found in each sample.
For example, at the function-level we substitute the first 10 variable tokens in a sample with the artificial tokens \tk{VAR0} - \tk{VAR9}, and substitute all the other ones with the abstract token \tk{VAR}. At the file-level, we retain the first 200 variables instead.
We also retain the concrete token for selected PHP functions such as \mmp$query, exec, strip_tags$ which are relevant to the vulnerabilities we study, and typically represent sinks or santizers.

Next, we turn each token into a number, using the \cmd{LabelEncoder} from \cmd{scikit-learn}\cite{LabelEncoder} which, given a vocabulary of tokens, maps each to a sequential natural number. 

The GRU that consumes our token sequences requires vectors of fixed length as inputs. 
For function-level samples we use a fixed length of 200, and for file-level samples, that tend to be significantly longer, we use 3000.
In each case, if a sample has fewer tokens we pad it with zeros, and if it has more tokens we keep the maximum number of tokens allowed, starting from the end of the token sequence.
This can lead to loss of information, but is a commonly accepted approach to analyse variable-length inputs, and in particular source code, with fixed-length neural architectures.

\subsubsection{CFG}

At the function-level, we parse each sample into an AST using \cmd{phply}. 
Then we extract the CFG, where each node represents a line of code, using our adaptation of code from \cmd{wirecaml}~\cite{Kronjee2018Wirecaml}.
We use the same procedure as for sequences of tokens above, but with a fixed length of 20, to turn each node into a numerical vector.
To transform the control flow graph into a tensor suitable for consumption by our GCN, 
we collate the nodes into a 2d matrix where each row contains the features of the node corresponding to the row index.
Next we represent the CFG edges as a vector of tuples $(i,j)$ representing a directed edge from node $i$ to node $j$. 

The file-level process is analogous, except that we use \cmd{joernphp}~\cite{Backes2017JoernPHP} to parse and extract the CFGs from the samples, as it proved to be more robust for large files.

%\vspace{-2mm}
\subsection{Model Architecture}
Figure~\ref{fig:arch} illustrates the overall architecture of \DT. We now describe each component and summarize the architectural parameters.

\subsubsection{Embedding Layer}

The role of the embedding layer is to transform each numerical input produced in the preprocessing stage into a vector of real numbers, encoding that input as a combination of factors in a lower-dimensional space.
We have two embedding layers; one for the token sequence and one for the CFG representation. Both embedding layers use vectors of length 100, which are learned via backpropagation during training and initialised at random. More formally, these layers are simply a mapping from a numerically tokenised function $t_i$, to a vector $v_i \in \mathbb{R}^{100}$.

\subsubsection{GRU}

We extract features from the sequence of tokens representations using a multi-layer bidirectional Gated Recurrent Unit~\cite{cho2014learning} which can learn long term dependencies between the tokens. 
Code patterns, such as those leading to vulnerabilities, heavily depend on the syntax of a programming language and the local context in which they appear. Most tokens carry information about the next token in the sequence. This information flow propagates until the end of a code statement. The GRU takes advantage of this information flows by learning bidirectional sequences  (i.e. forwards and backwards) of code tokens throughout the source code.

Internally, each layer of the GRU computes the following function for each element in the input sequence:

\begin{equation*}
        \begin{array}{ll}
            r_t = \sigma(W_{ir} x_t + b_{ir} + W_{hr} h_{(t-1)} + b_{hr}) \\
            z_t = \sigma(W_{iz} x_t + b_{iz} + W_{hz} h_{(t-1)} + b_{hz}) \\
            n_t = \tanh(W_{in} x_t + b_{in} + r_t * (W_{hn} h_{(t-1)}+ b_{hn})) \\
            h_t = (1 - z_t) * n_t + z_t * h_{(t-1)}
        \end{array}
\end{equation*}
where $h_t$ is the hidden state at time $t$, $x_t$ is the input at time $t$, $h_{(t-1)}$ is the hidden state of the layer at time $t-1$ or the initial hidden state at time $0$, and $r_t$, $z_t$, $n_t$ are the reset, update, and new gates, respectively. $\sigma$ is the sigmoid function, and $*$ is the Hadamard product. 
Since we are implementing a multilayer GRU, the input $x^{(l)}_t$ of the $l$-th layer ($l >= 2$) is the hidden state $h^{(l-1)}_t$ of the previous layer multiplied by dropout, $\delta^{(l-1)}_t$ where each $\delta^{(l-1)}_t$ is a Bernoulli random variable which is $0$ with probability dropout \cite{Pytorch2019GRU}.
The output we take from the GRU is the concatenation of the hidden states at the beginning and end of each layer.

\subsubsection{GCN}

The CFG represents the control dependencies of functions and statements in a code sample. These approximate the flow of information from untrusted sources to sensitive sinks typical of injection vulnerabilities. Therefore, we extract features from the CFG using a Graph Convolutional Network~\cite{kipf2016semi}, which is able to embed such dependencies into our model, and learn their significance via backpropagation.

Internally we use three layers of a GCN followed by Edge Pooling. 
Let $\mathbf{X}$ be a graph node vector, and $\mathbf{\hat{A}} = \mathbf{A} + \mathbf{I}$ the adjacency matrix of the graph, with inserted self-loops. 
The equation
\begin{equation*}
        \mathbf{X}^{\prime} = \mathbf{\hat{D}}^{-1/2} \mathbf{\hat{A}}
        \mathbf{\hat{D}}^{-1/2} \mathbf{X} \mathbf{\Theta}
\end{equation*}
defines the convolved signal matrix $\mathbf{X'}$, where $\hat{D}_{ii} = \sum_{j=0} \hat{A}_{ij}$ denotes the diagonal degree matrix and $\mathbf{\Theta}$ denotes the convolutional filter parameters~\cite{Fey2019GCNConv}.

\subsubsection{Classification}

We take the output of the graph convolutional layers and flatten it using max pooling. 
The output of the graph convolutional layers are node vectors of length 4000.
The max pooling scans the ith element of each node and selects the maximum values as the ith element of the output vector. Mathematically:
\begin{equation*}
    o_i = max_{n=1}^{N} x_{in}
    %\vspace{-0.1mm}
\end{equation*}
where $o_i$ is the ith element of the output vector, $x_in$ is the ith element of the nth node in the output graph, and $N$ is the number of nodes in the graph.

We combine the output vector of the GCN with the output vector of the GRU and feed them to the linear classification layers. We have 3 linear classification layers, each with a dropout of 0.3 to combat overfitting, followed by a ReLU activation function. The final output of the ReLU is a probability vector of length 4, representing the confidence of assigning the sample to each class.

\subsubsection{Architectural Parameters} 
We have tested alternative hyper-parameters settings to tune the model, and we found that the current configuration provides us with the best detection performance across different types of vulnerability. 
Table \ref{tab:deeptective_arch} shows the details of DeepTective architecture by layers.

\begin{table}[h]
    \scriptsize
%    \captionsetup{font=footnotesize}
    \caption{\DT\ architectural parameters}
    \label{tab:deeptective_arch}
    \centering
    \begin{tabular}{p{1.8cm} p{0.5cm} p{1.1cm} p{0.8cm} p{0.9cm} p{0.8cm}}
    \toprule
    \textbf{Layer} & \textbf{Output size} & \textbf{Edge-pool size} & \textbf{Dimension} & \textbf{Activation function} & \textbf{Dropout}\\
    \midrule
    Embedding & 200 & None  & 100 & None & None\\
    GRU & 200 & None & None & None & None\\
    GRU & 200 & None & None & None & None\\
    GRU & 200 & None & None & None & None\\
    Embedding & 2000 & None & 100 & None & None\\
    GCNConv & 2000 & 2000 & None & ReLU & None\\
    GCNConv & 4000 & 4000 & None & ReLU & None\\
    GCNConv & 4000 & 4000 & None & ReLU & None\\
    Fully connected & 1000 & None & None & ReLU & 0.3\\
    Fully connected & 500 & None & None & ReLU & 0.3\\
    Fully connected & 4 & None & None & ReLU & None\\
    \bottomrule
    \end{tabular}
    %\vspace{-6mm}
\end{table}

\section{Datasets}\label{sec:dataset}

In order to evaluate a supervised vulnerability detection model, we need to build datasets with vulnerable and non vulnerable samples.
We are interested in two kind of samples: entire files, or individual functions.
We label the samples as Safe, XSS, SQLi and OSCI, where the latter 3 labels together are the Unsafe ``virtual'' label.
We extract the samples from synthetic data (SARD) and real-world projects (GitHub), as detailed below.
In order to support further research in the area, and facilitate the comparison between different approaches, we plan to make our datasets available to the public.

\subsection{Synthetic Samples}
\newcommand\sard{SARD$^\#$}
\newcommand\sardo{SARD}

The Software Assurance Reference Dataset project \cite{sard} is a collection of code samples for multiple programming languages. The objective is to enable researchers and developers to evaluate alternative methods for detecting different types of bugs. 
Below, we consider the subset of SARD for PHP vulnerabilities~\cite{sardphp}.
Each sample is a short standalone file with no external dependencies. Samples are generated by a tool called the PHP Vulnerability Test Suite Generator \cite{sardphpgenerator}. 
The dataset contains both safe and unsafe samples for different vulnerability types.
A separate metadata file lists the line considered responsible for the vulnerability of each unsafe sample.

\begin{figure}[!t]
    %\vspace{-2mm}
%    \captionsetup{font=footnotesize}
   \label{fig:SARD}
    \begin{center}
\hrule
\begin{lstlisting}[language=PHP,basicstyle=\fontsize{8pt}{8pt}\selectfont\ttfamily]
<!DOCTYPE html><html>
<head><style><?php
$array = array();
$array[] = 'safe' ;
$array[] = $_GET['userData'] ;
$array[] = 'safe' ;
$tainted = $array[1] ;
$tainted = http_build_query($tainted);
//flaw
echo $tainted ;
?></style></head>
<body><h1>Hello World!</h1></body>
</html>
\end{lstlisting}
\hrule
    \end{center}
\caption{XSS test case from the SARD dataset.}
\end{figure}

Consider the example of a web page vulnerable to XSS from SARD reported in Figure~\ref{fig:SARD}.
We can immediately see that SARD samples are rather simplistic and unlikely to reflect code in real world projects (as in Table~\ref{tab:code_snippet}), so it is not clear if models learned on SARD can be transferred to other code bases. 
Despite that, SARD has been widely used in previous studies of vulnerabilities, including specifically for PHP~\cite{Kronjee2018Wirecaml,Fang2019Tap}.
The advantages of SARD are that vulnerabilities are guaranteed to be self-contained in the samples, and each sample has very few irrelevant lines of code.
This helps focusing the learning process.
Besides, the labels of SARD samples are highly accurate, which is a primary concern for supervised machine learning approaches.

We extract the PHP code from each SARD safe and unsafe sample for XSS, SQLi and OSCI.
Some SARD samples are very similar to each other. For example, there is a variant of the listing above where the \cmd{<style>} tags are replaced with \cmd{<script>} tags. This introduces duplicates in our code-only dataset, which we remove.
We only collect file-level samples for SARD as each sample is already very short, and any function, if present, only contains very limited code (typically 1 line).
We denote by \sard\ our derived dataset.

The number of samples in the original SARD dataset and in our dataset are reported in Table~\ref{tab:alldata}.

\subsection{Realistic Samples}

Besides the focused, synthetic samples from \sard, we want to collect a dataset representing vulnerabilities as they actually appear in realistic PHP projects. 

GitHub hosts source code for PHP projects of all sizes, ranging from the extremely popular WordPress framework to a beginner's first PHP snippet.
In order to select representative vulnerabilities, we searched the National Vulnerability Database (NVD)~\cite{nvd} for CVE entries labelled with the CWE identifier of XSS (CWE-79), SQLi (CWE-89) and OSCI (CWE-78). 
We extracted from the references of each relevant CVE any GitHub commit URL, and cloned the corresponding PHP repositories.
In combination, we also cloned from GitHub some of the largest and most commonly used open source PHP projects: \cmd{Moodle}, \cmd{CodeIgniter}, \cmd{Drupal}, \cmd{ILIAS}, \cmd{phppmyadmin}, \cmd{wikia}, \cmd{magento2}, \cmd{simplesamlphp} and \cmd{WordPress}.

\subsubsection{Sample Extraction}
We search the commit history of each cloned project for keywords related to the vulnerabilities we are interested in, including ``xss'', ``sqli'' and several variants.
There are a few commit messages that report fixing both XSS and SQLi vulnerabilities: we exclude these, as multi-label classification is beyond the scope of this project. 
When we come across a relevant commit, we extract the vulnerable version of the affected files, and add to each file the label for the corresponding vulnerability.
These constitute our file-level positives.
From the same version of the repository, we save the files not affected by the commit as our file-level negatives.

To build a function-level dataset, we make the assumption that the presence in the function body of patched lines from a relevant commit implies that a function is vulnerable.
We take a vulnerable file as identified above, start from the lines changed by relevant commits, and use interval trees to extract from the source code the functions containing the changed lines.
These constitute our function-level positives.
We also extract from each file the functions that do not contain any line changed in the commit, and save them as our function-level negatives.

\subsubsection{Label Noise}
The approach described above may introduce noise in the labelling of samples. 
Files may be mislabelled when a commit message misidentifies a vulnerability.
Vulnerable files with a commit message that does not mention a vulnerability fix, and files which contain vulnerabilities not known or fixed by the developers, will be mistakenly labelled as negatives.
A vulnerability-relevant commit may also include unrelated changes to non-vulnerable files. 
These files will be mistakenly labelled as positives. 
Similarly, if a vulnerable file contains changes for both vulnerable and non-vulnerable functions, the latter will be mislabelled as positives.
To limit these effects, we ignore commits modifying more than 20 files, and we discard changes that only consist in deleting lines of code, as both cases are mostly associated with code refactoring.
We manually inspected 10\% of the files labelled as positives, and did not detect any mislabelling.

\subsubsection{Datasets}
We denote the file-level dataset by \gitfi, and the function-level dataset by by \gitfu. The number of samples of each class in \gitfi\ and \gitfu\ are reported in Table~\ref{tab:alldata}.
Note that the relation between the number of samples in the two datasets is not straightforward. \gitfu\ has more negatives samples, as a file with a vulnerable function may contain a sizeable number of functions not affected by the commit. On the other hand, \gitfi\ has more positives because some vulnerabilities are not located inside function bodies. % SM: if not, why are there more?

\begin{table}[!t]
    %\vspace{-2mm}
%    \captionsetup{font=footnotesize}
    \caption{Number of samples in relevant datasets.}
   \label{tab:alldata}
    \begin{center}
    \scriptsize
    \begin{tabular}{l l l l l }
    \toprule
                 \textbf{Dataset} &    \textbf{Safe}&\textbf{XSS}  & \textbf{SQLi} & \textbf{OSCI}\\
\midrule
    \sardo &  16240 & 4352 &912 & 624 \\
    \sard &2928 &960 & 288&250 \\
    \gitfi & 2726 & 2117 & 604 & 7\\ 
    \gitfu & 4288 & 726 & 428 & 11\\ 
    \bottomrule
    \end{tabular}
    \end{center}
    %\vspace{-7mm}
\end{table}

\section{Model Evaluation}

We evaluate \DT\ on the separate tasks of function classification and file classification.
For each task, we train and test the model on data from SARD, GitHub, and from both. 
This allows us to compare the difference between using synthetic and real-world samples. 
Furthermore, we compare the classification performance of \DT\ with previous work, and identify interesting variations between the approaches.

\subsection{Methodology}%\label{sec:classification}

\subsubsection{Experimental Setup}

For both experiments, we use Pytorch 1.5 and Torch Geometric 1.5.0 with CUDA 10.1 on top of Python 3.8.1. We train the model on a computer running Intel Xeon Skylake CPU (40 cores), 128GB RAM and Nvidia GTX Titan XP. We use Weights \& Biases \cite{wandb} as our main experimental management tool to track the each run throughout this work.

\subsubsection{Performance Criteria}

For each experiment, we report true negatives (TN), false negatives (FN), true positives (TP), false positives (FP), accuracy, precision, recall and F1-score.

Accuracy measures the percentage of correctly predicted samples, but is not very significant when test classes are imbalanced, like in the case of vulnerabilities which are very rare compared to safe samples. A trivial classifier marking everything as safe would have very high accuracy in the real world, but little use.
Precision measures how many of the reported vulnerabilities are actual vulnerabilities: it tells us if it is worth investigating the results of the classifier.
Recall measures the percentage of existing vulnerabilities that the classifier is able to discover: it tells us how worried we should still be after running the tool.
Finally, the F1 score summarises numerically the balance between precision and recall. 

Note that in Tables~\ref{tab:exp_funclevel} and \ref{tab:exp_filelevel} we report only the figures for the binary classification problem where the positives classes XSS, SQLi and OSCI are merged in the Unsafe class.
This is to simplify exposition, and because ultimately we care mostly about detecting vulnerabilities, irrespective of their specific label.
Internally though we train our model and measure results for multiclass-classification, and will report relevant details where appropriate.

\subsubsection{Model Training}

Since this is a multiclass-classification problem, we use cross-entropy as our loss function.
The training process uses a batch size of 64 along with an Adam optimiser and a learning rate of $10^{-5}$. Alongside this, we implement a learning rate scheduler that reduces the learning rate if the loss plateaus. Lastly, we split the dataset for training/validation/test to 80/10/10, and stratify data according to their classes. With the model and hyper-parameters in place, we train the model for 150 epochs to maximise the learning potential of our model.

\subsection{Classification}\label{sec:classification}

We perform two experiments to investigate different learning patterns across function-level and file-level granularity. Different granularity levels are expected to behave differently based on the information provided for the model to learn.

\subsubsection{Function-Level Granularity}

Table \ref{tab:exp_funclevel} shows the result of testing our function-level model on the \sard, \gitfu\ and combined dataset \combofu, after training on each of them respectively. 

\begin{table}[!t]
    %\vspace{-1.3mm}
    \scriptsize
%    \captionsetup{font=footnotesize}
    \caption{Function-level granularity results.}
    \label{tab:exp_funclevel}
    \centering
    \begin{tabular}{p{0.8cm} p{0.7cm} p{0.2cm} p{0.2cm} p{0.2cm} p{0.2cm} p{0.7cm} p{0.7cm} p{0.5cm} p{0.4cm}}
    \toprule
    \textbf{Model}& \textbf{Testing  \hspace{0.2cm}set} & \textbf{TN} & \textbf{FN} & \textbf{TP} & \textbf{FP} & \textbf{Accuracy (\%)} & \textbf{Precision (\%)} & \textbf{Recall (\%)} & \textbf{F1 (\%)} \\
    \midrule
    Func-S& \sard & 277 & 0 & 149 & 16 & 96.38 & 90.30 & 100.0 & 94.90\\
    (\sard) & \gitfu & 4182 & 1133 & 42 & 105 & 76.88 & 28.57 & 3.57 & 6.35\\
    & \combofu & 4455 & 1134 & 190 & 105 & 78.27 & 60.32 & 14.35 & 23.18\\
    \midrule
    Func-G& \sard & 2361 & 1248 & 240 & 567 & 54.46 & 29.74 & 16.13 & 20.92\\
    (\gitfu) & \gitfu & 366 & 65 & 53 & 63 & 75.14 & 45.69 & 44.92 & 45.30\\
    & \combofu & 2727 & 1313 & 293 & 630 & 56.74 & 31.74 & 18.24 & 23.17\\
    \midrule
    Func-A& \sard & 290 & 0 & 149 & 3 & 99.32 & 98.03 & 100.0 & 99.0\\
    (\combofu) & \gitfu & 378 & 72 & 46 & 51 & 76.05 & 47.42 & 38.98 & 42.73\\
     & \combofu & 661 & 70 & 197 & 61 & 85.84 & 76.36 & 73.78 & 75.05\\
    \bottomrule
    \end{tabular}
    %\vspace{-3mm}
\end{table}

\stitle{Func-S}
The Func-S model is trained on the \sard\ dataset. It achieves great performance when testing on the \sard\ dataset itself (despite training data not overlapping with testing data). The precision is 90.30\% with a recall of 100\% and F1 of 94.9\%. The perfect recall score means that all vulnerable PHP samples are correctly classified as true positives. However, Func-S fails spectacularly on the real-world \gitfu\ dataset, with precision and recall down respectively to 28.57\% and 3.57\%. We hypothesize that this failure to generalise is due to the highly skewed and homogeneous nature of \sard\ samples on which the model is trained. In particular, the model fails to detect most of the vulnerable \gitfu\ samples (1133 FN). On inspection, \sard\ vulnerable samples are short and focused around the vulnerability, whereas \gitfu\ vulnerable functions may contain a lot of irrelevant context, and more varied vulnerability patters. As can be expected, the performance on \combofu\ is roughly a weighted average of the preceding two.

\stitle{Func-G}
The results for the Func-G model are qualitatively similar, but the performance on the same distribution (the \gitfu dataset) is still disappointing in absolute terms, with 45.69\% precision, 44.92\% recall and 45.29\% F1. We believe this shows that the function-level model is not appropriate for real world code. In fact, by manually inspecting \gitfu\ samples we can observe that although a vulnerability may in effect be present inside a function, the vulnerable line by itself is not sufficient to detect the function as vulnerable. As an extreme example, the identify function can be considered as a vulnerable instance of a function to sanitize user input: but inspecting the identity function by itself gives no clues to the presence of a vulnerability. This observation motivated us to explore file-level granularity, but first we investigate if combining \sard\ and \gitfu\ could introduce synergies which improve the classification performance.

\stitle{Func-A}
The results for the Func-A model show a noticeable improvement on \sard\ over the already high performance of Func-S. On the other hand, the performance on \gitfu\ is similar but slightly worse (F1 score) than the one of Func-G, so the benefit of training across datasets was only felt in one direction. Finally, note that the jump in performance on \combofu\ is mostly an artefact of the lower number of samples available for testing, as 90\% of both \sard\ and \gitfu\ data is used for training. This leads to a higher weight given to the \sard\ performance in comparison to the Func-S case.

\begin{figure*}[!t]
%\vspace{-3mm}
    
\begin{tabular}{cc}
    \includegraphics[scale=0.75]{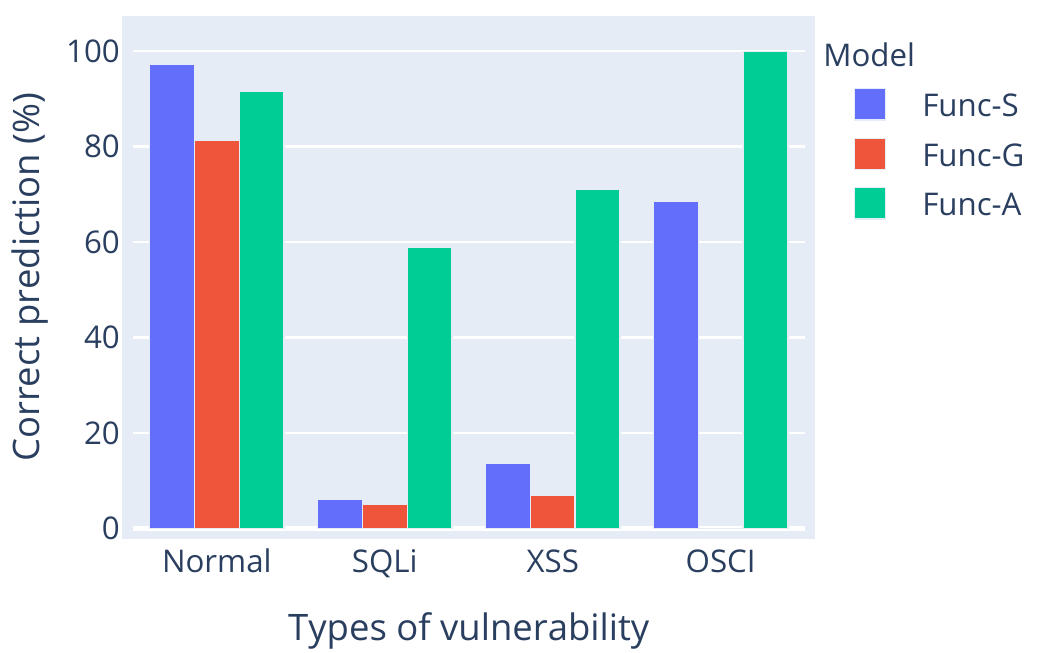} & \includegraphics[scale=0.75]{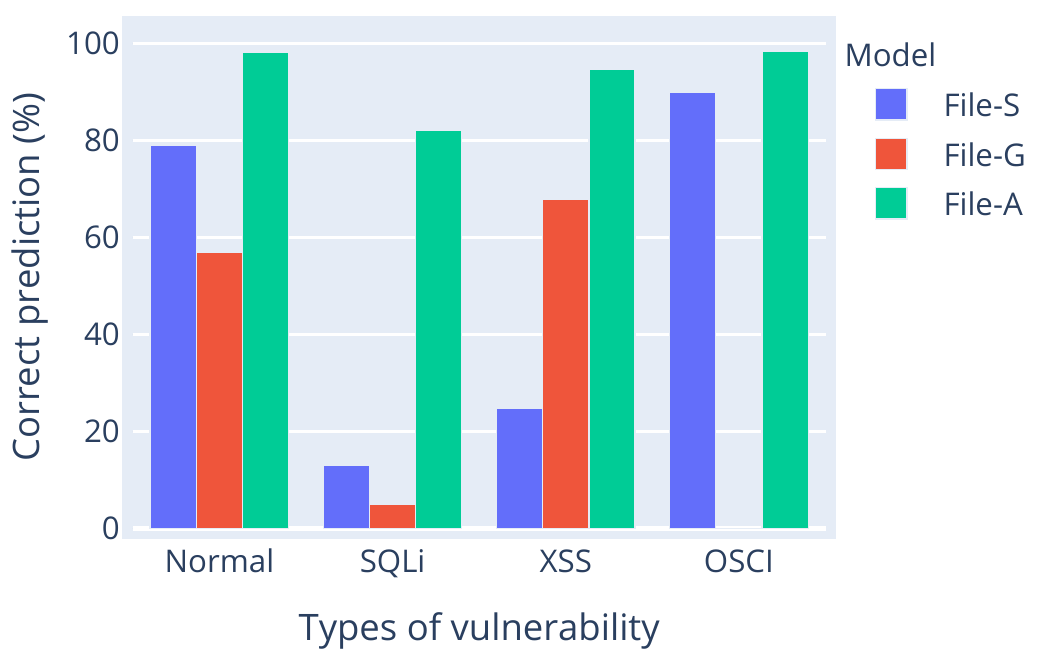}\\
    (A) Function-level & (B) File-level 
\end{tabular}
    \caption{Distribution of correctly predicted samples across different types of vulnerability.}
    \label{fig:dist_prediction_func}\label{fig:dist_prediction_file}
    \centering
    %\vspace{-2mm}
\end{figure*}

Figure \ref{fig:dist_prediction_func}(A) compares the percentage of correct predictions for each fine-grained class on the \combofu\ test set, for Func-S,-G and -A.
The main observation is that the Func-A model shows a reliable predictive capability, with all classes above 60\%. Overall, at the function-level, combining datasets from different data distributions allows the model to learn more vulnerability patterns, which help the model to generalise its detection ability across different code writing styles and application domains.

\subsubsection{File-Level Granularity}

We now want to test if providing more context to the model improves its ability to learn vulnerable code patterns. Contextual information is made available to the model by switching from function-level to file-level granularity of samples, and adapting the model as described in Section~\ref{sec:dt}. 
Table \ref{tab:exp_filelevel} shows the result of training our model at the file-level granularity and testing on \sard, \gitfi\ and their combination \combofi.

\begin{table}[!t]
    \scriptsize
%    \captionsetup{font=footnotesize}
    \caption{File-level granularity results.}
    \label{tab:exp_filelevel}
    \centering
    \begin{tabular}{p{0.74cm} p{0.7cm} p{0.2cm} p{0.2cm} p{0.2cm} p{0.2cm} p{0.7cm} p{0.7cm} p{0.5cm} p{0.4cm}}
    \toprule
    \textbf{Model}& \textbf{Testing  \hspace{0.2cm}set} & \textbf{TN} & \textbf{FN} & \textbf{TP} & \textbf{FP} & \textbf{Accuracy (\%)} & \textbf{Precision (\%)} & \textbf{Recall (\%)} & \textbf{F1 (\%)} \\
    \midrule
    File-S& \sard & 1624 & 0 & 589 & 0 & 100 & 100.0 & 100.0 & 100.0\\
    (\sard) & \gitfi & 1817 & 2263 & 465 & 909 & 36.89 & 33.84 & 17.05 & 22.67\\
    & \combofi & 3439 & 2263 & 1054 & 911 & 55.13 & 53.64 & 31.78 & 39.91\\
    \midrule
    File-G& \sard & 9143 & 2010 & 3878 & 7097 & 54.62 & 35.33 & 65.86 & 45.99\\
    (\gitfi) & \gitfi & 251 & 44 & 229 & 22 & 83.33 & 91.24 & 83.88 & 87.40\\
    & \combofi & 9396 & 2054 & 4107 & 7117 & 55.32 & 36.59 & 66.66 & 47.25\\
    \midrule
    File-A& \sard & 1624 & 1 & 588 & 0 & 99.95 & 100.0 & 99.83 & 99.92\\
    (\combofi) & \gitfi & 240 & 32 & 241 & 33 & 82.78 & 87.96 & 88.28 & 88.12\\
    & \combofi & 1864 & 34 & 828 & 33 & 96.56 & 96.17 & 96.06 & 96.11\\
    \bottomrule
    \end{tabular}
    %\vspace{-4mm}
\end{table}

\stitle{File-S}
The File-S model achieves perfect scores for all the metrics when testing on the \sard\ dataset itself.
The improvement over the performance of Fun-S, which is trained on the same dataset, must then be due entirely to the model adaptations for file-level granularity, including a larger vocabulary size.
The results of \gitfi\ are also better than Fun-S but still not practically useful.

\stitle{File-G}
A substantial improvement instead is observed on the performance of the File-G model, in particular on \gitfi.
It achieves 91.24\% precision and 83.88\% recall, for an F1 score of 87.40\%. 
This shows that although \gitfi\ is a harder dataset to learn, consisting of highly diverse PHP files from popular projects, file-level granularity provides enough of the missing information to achieve usable performance.

\stitle{File-A}
Training on the combined dataset has the effect of slightly reducing the perfect performance of File-S on \sard, but yields a larger increases over the F1 score of File-G on \gitfi.
In particular File-A finds more real-world vulnerabilities (increase of TP) but at the price of a few more false alarms (increase of FP).

\iffalse
\begin{figure}[h]
    %\vspace{-3mm}
    \scriptsize
    \captionsetup{font=footnotesize}
    \includegraphics[scale=0.6]{img/dist_prediction_filelevel.pdf}
    \captionsetup{justification=centering, margin=0pt}
    \caption{Distribution of correctly predicted samples across different types of vulnerability for all file-level models}
    \label{fig:dist_prediction_file}
    \centering
    %\vspace{-2mm}
\end{figure}
\fi

Figure \ref{fig:dist_prediction_file}(B) compares the percentage of correct predictions for each fine-grained class on the \combofi\ test set, for File-S,-G and -A.
The average predictive capability of File-A is higher than 80\% for all classes, hence the additional contextual information provided at the file-level has a significant impact also at the multi-class classification level.

\subsection{Tool Comparison}\label{sec:tools}

We compared the classification performance of \DT\ File-A, our best model, with selected publicly available tools to find PHP vulnerabilities, based on machine learning (\cmd{wirecaml} and \cmd{TAP}) or static analysis (\cmd{progpilot}, \cmd{RIPS} and \cmd{WAP})~\cite{Kronjee2018Wirecaml,Fang2019Tap,progpilot,Dahse2010rips,Medeiros2014WAP}.
We ran all the tools above on the same test sets from the \sard\ and \gitfi\ datasets which we used in Section~\ref{sec:classification} to evaluate File-A. We measured the tools detection performance, which is reported in Table~\ref{tab:compare_sard}. Note that \cmd{wirecaml} is made of two binary classifiers for XSS and SQLi and thus we report the performance of each individual classifier. Furthermore, vulnerabilities of the class that is not being classified by a \cmd{wirecaml} classifier were deemed as safe samples when judging performance.
Machine learning tools often perform better when trained and tuned using their authors' datasets.
Hence, we used \cmd{wirecaml} and \cmd{TAP} trained on their respective datasets, effectively testing their ability to generalise to new datasets. 

\begin{table}[!t]
    \scriptsize
%    \captionsetup{font=footnotesize}
    \caption{Comparison: \DT\ File-A vs. selected tools.}
    \label{tab:compare_sard}
    \centering
    \begin{tabular}{p{1.7cm} p{0.2cm} p{0.2cm} p{0.2cm} p{0.2cm} p{0.7cm} p{0.7cm} p{0.5cm} p{0.4cm}}
    \toprule
    \textbf{Tool name} & \textbf{TN} & \textbf{FN} & \textbf{TP} & \textbf{FP} & \textbf{Accuracy (\%)} & \textbf{Precision (\%)} & \textbf{Recall (\%)} & \textbf{F1 (\%)} \\
    \toprule
    \multicolumn{9}{c}{\textbf{A}: Results for \sard\ dataset.}\\
    \toprule
    DeepTective & 1624 & 1 & 588 & 0 & \textbf{99.95} & \textbf{100.0} & 99.83 & \textbf{99.92}\\
    \midrule
    TAP & 1584 & 96 & 493 & 40 & 93.85 & 92.50 & 83.70 & 87.88\\
    wirecaml-XSS & 470 & 50 & 385 & 1308 & 38.64 & 22.74 & 88.51 & 36.18\\
    wirecaml-SQLi & 1496 & 0 & 91 & 626 & 71.71 & 12.69 & \textbf{100.00} & 22.52\\
    \midrule
    progpilot & 629 & 304 & 285 & 995 & 41.30 & 22.27 & 48.39 & 30.50\\
    WAP & 1342 & 477 & 112 & 282 & 65.70 & 28.43 & 19.02 & 22.79\\
    RIPS & 1440 & 497 & 92 & 184 & 69.23 & 33.33 & 15.62 & 21.27\\
    \toprule
    \multicolumn{9}{c}{\textbf{B}: Results for \gitfi\ dataset.}\\
    \toprule
    DeepTective & 240 & 32 & 241 & 33 & 82.78 & \textbf{87.96} & \textbf{88.28} & \textbf{88.12}\\
    \midrule
    TAP & 233 & 262 & 11 & 40 & 44.69 & 21.57 & 4.03 & 6.79\\
    wirecaml-XSS & 299 & 171 & 41 & 35 & 62.27 & 53.95 & 19.34 & 28.47\\
    wirecaml-SQLi & 484 & 60 & 0 & 2 & \textbf{88.64} & 0.00 & 0.00 & 0.00\\
    \midrule
    progpilot & 265 & 257 & 16 & 8 & 51.47 & 66.67 & 5.86 & 10.77\\
    WAP & 160 & 154 & 119 & 113 & 51.10 & 51.29 & 43.59 & 47.13\\
    RIPS & 256 & 225 & 48 & 17 & 55.68 & 73.85 & 17.58 & 28.40\\
    \bottomrule
    \end{tabular}
    %\vspace{-3mm}
\end{table}

The results show that \DT\ significantly outperformed the other tools in terms of F1 score. 
\cmd{TAP} achieved a high F1 on the synthetic \sard\ dataset, but showed poor performance on the realistic samples from \gitfi.
\cmd{wirecaml-SQLi} achieved a high accuracy on \gitfi, but at the price of null precision and recall. Note that the same tool had perfect recall on the \sard\ dataset.
On a synthetic dataset intersecting with with our \sard, \cite{Fang2019Tap} measured F1 scores of 98.8\% and 97.5\% for  \cmd{TAP} and \cmd{wirecaml} respectively. Our failure to replicate a similar result for those (pre-trained) tools on \sard\ points to the difficulty for some machine learning models to generalise even to related datasets.
We have noted above how a perfect 100\% F1 for File-S on \sard\ translated into a poor 22.67\% F1 for the same model on \gitfi.
That result is in line with the drop observed in the performance of all the tools above from testing on \sard\ to testing on \gitfi. 
We believe our results show that evaluating tools only on synthetic datasets is not a sufficient guarantee of practical performance, and that \DT\ File-A stands out in its ability to perform well on realistic samples.

\section{Practical Experiments}

In order to evaluate the practical usefulness of our model, we ran it on a number of PHP projects which we did not include in our \gitfi\ dataset.
In particular we want to estimate the execution performance, to ensure that the tool can scale also to large projects, and assess it usability for actual vulnerability detection.

For these experiments we have chosen 13 software projects divided in two sets: 8 \textbf{popular projects} and 5 \textbf{smaller plugins}.
The popular projects are listed in the top 50 GitHub repositories (based on stars), that use PHP as their primary language, and span from a few hundred kilobytes to tens of megabytes in size. These are meant to be a representative benchmark for the execution performance. We expect the popular projects to be carefully reviewed, hence we make the assumption that they currently have no security vulnerabilities, and we assume no TP and FN for classification purposes.
We also collect 5 WordPress plugins projects, with a limited number of users (less than 20,000), to increase the likelihood of them containing an undiscovered security vulnerability. Projects with a limited user base may have a smaller development team lacking security expertise, or be subject to less scrutiny than popular projects.

Below we report the execution performance and accuracy for both sets, then we dig deeper on the smaller plugins sets to hunt for vulnerabilities, to limit the effort necessary in manually reviewing positives.

\subsection{Execution Performance}

The size of the software projects considered varies from 110KB with 2713 lines of codes (LoC) to 27MB with 242,299 lines of codes. The size and LoC distribution of these software projects reflect the distribution of real-world projects as some projects are small and large in scale.

% Definition of proc time, inf time and accuracy.

To evaluate the execution performance of \DT\ across real-word software projects, we use the following performance metrics:

\begin{itemize}
    \item\stitle{Lines of codes (LoC)} The number of lines of codes in each file for all the PHP files in a specific software project.
    \item\stitle{Processing time} The time taken (in seconds) to process and transform a PHP file to the data structure used by our detection model. This process includes the creation of token sequences and CFGs.
    \item\stitle{Inference time} The time take to perform the classification of all the PHP files in a specific software project.
    \item\stitle{Time/LoC} The average total time taken (processing and inference) per line of code for a software project.
    \item\stitle{Time/File} The average total time taken (processing and inference) per file in a software project.
\end{itemize}

% Table from the File-G (partial) model

\begin{table*}[t]
    \scriptsize
%    \captionsetup{font=footnotesize}
    \caption{\DT\ execution performance.}
    \label{tab:execution_performance}
    \centering
    \begin{tabular}{p{3.2cm} p{0.9cm} p{0.5cm} p{0.6cm} p{1.1cm} p{0.9cm} p{1cm} p{1cm} p{1.5cm} p{1.6cm}}
    \toprule
    \textbf{Software project}& \textbf{Size (bytes)} & \textbf{PHP files} & \textbf{LoC} & \textbf{Processing time (s)} & \textbf{Inference time (s)} & \textbf{Time (s)/LoC} & \textbf{Time (s)/File} & \textbf{File-A accuracy(\%)} & \textbf{File-G accuracy(\%)}\\
    \midrule
    \multicolumn{10}{c}{\textbf{A}: Results for popular projects}\\
    \midrule
    Codeigniter & 7,416,704 & 669 & 138495 & 728.4836 & 7.4599 & 0.00531 & 1.10006 & 53.81 & 56.35 \\
    Composer & 2,342,547 & 252 & 53518 & 384.9617 & 3.1456 & 0.00725 & 1.54011 & 55.16 & 72.22 \\
    Grav & 5,955,146 & 347 & 60922 & 400.0879 & 4.0205 & 0.00663 & 1.16458  & 55.62 & 62.25\\
    Guzzle & 352,741 & 32 & 4555 & 28.1737 & 0.7210 & 0.00634 & 0.90296 & 50.00 & 59.38\\
    Laravel & 110,595 & 53 & 2713 & 17.1215 & 0.8413 & 0.00662 & 0.33892 & 69.81 & 75.47\\
    PHPMailer & 381,439 & 55 & 2185 & 20.6959 & 0.9937 & 0.00993 & 0.39436 & 96.36 & 96.36\\
    PHPUnit & 1,373,437 & 323 & 35367 & 225.5044 & 3.8504 & 0.00648 & 0.71008 & 66.80 & 83.59\\
    Symphony & 27,052,061 & 2676 & 242299 & 1699.9081 & 26.2258 & 0.00712 & 0.64504 & 75.85 & 75.67\\
    \midrule
    \multicolumn{10}{c}{\textbf{B}: Results for smaller plugins}\\
    \midrule
    Appointment Booking Calendar & 2,826,657 & 16 & 4735 & 50.1272 & 0.8017 & 0.01076 & 3.18306 & 31.25 & 56.25\\
    Payment Form for PayPal Pro & 1,005,490 & 13 & 4379 & 44.5420 & 0.7712 & 0.01035 & 3.48563 & 15.38 & 53.85\\
    PayPal for Digital Goods & 149,137 & 7 & 1152 & 6.1942 & 0.5617 & 0.00586 & 0.96514 & 57.14 & 42.86\\
    Sportspress & 4,834,097 & 256 & 50461 & 419.3428 & 3.4818 & 0.00838 & 1.65166 & 50.39  & 48.05\\
    Simple Jobs Board & 9,783,895 & 198 & 19775 & 108.8408 & 2.3262 & 0.00562 & 0.56145 & 86.87  & 86.36\\
    \midrule
    Total & 63,583,946 & 4897 & 620556 & 4133.9838 & 55.2009 & 0.00675 & 0.85546 & 61.98  & 71.37\\
    \bottomrule
    \end{tabular}
    %\vspace{-4mm}
\end{table*}

Table \ref{tab:execution_performance}-A shows the execution performance for popular projects. Symphony has the longest processing time of 1699.91 seconds as it has the most number of PHP files and LoC. Laravel has the shortest processing time of 17.12 seconds, despite having a higher number of LoC (2713) than PHPMailer (2185). This is due to the simpler structure of Laravel code, which is a lightweight PHP framework containing the wireframe to develop a PHP web application. In terms of inference time, the data shows a consistent trend based on the number of PHP files in a project. The higher the number of PHP files, the longer the time it takes to perform inference, as the process is done on the file-level granularity. Time/LoC metric demonstrates minor differences across all the software projects in GitHub. However, the Time/File metric shows some surprising pattern as Composer has the highest execution time per file even though the number of total PHP files and execution time are lower than other larger projects like Symphony and CodeIgniter. Composer \cite{Khliupko2017} is a dependency management tool for PHP projects, which allows the user to declare, update and manage external libraries. Based on this, it shows that the complexity of Composer contributes to the high Time/File performance metric as compared to other GitHub projects. 

Table \ref{tab:execution_performance}-B shows the execution performance for smaller plugins obtained from WordPress plugins website. The LoC for each project is consistent based on both the project size and the total number of PHP files. In terms of processing time and inference time, Sportspress takes much longer with 419.34 seconds and 3.48 seconds respectively, even though having fewer LoC than Simple Jobs Board. However, it is worth noting that Sportspress has a higher number of PHP files, and this significantly affects the execution time as the evaluation is done based on the file-level granularity. Surprisingly, in terms of Time/LoC and Time/File metrics, smaller projects like Appointment Booking Calendar and Payment Form for PayPal Pro recorded higher values as compared to larger projects like Sportspress and Simple Jobs Board. As for the popular projects, this variance reflects the different code complexity and style across different projects.

\begin{figure}[h]
  \scriptsize
 \centering
 \includegraphics[scale=0.55]{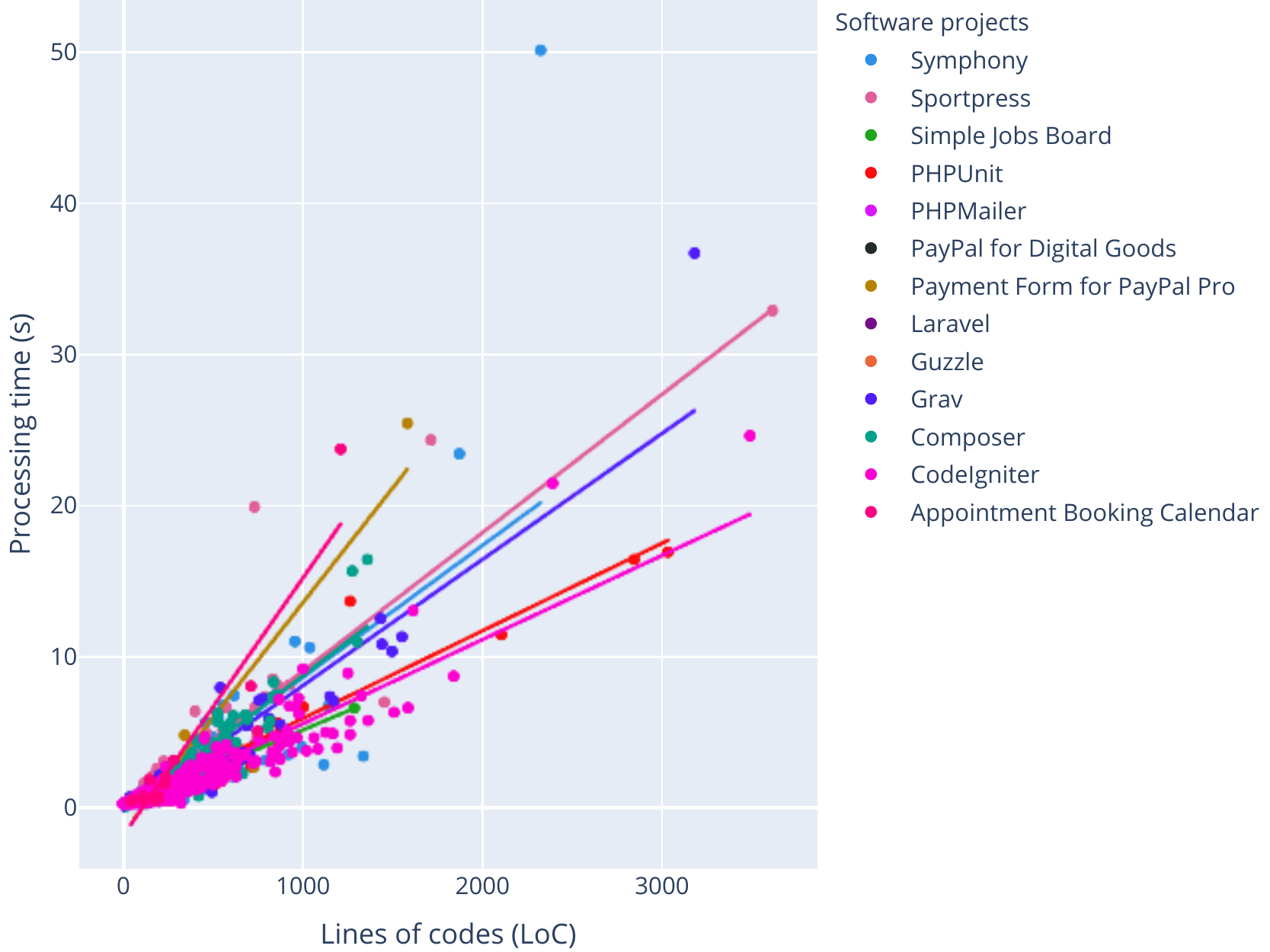}
    %\captionsetup{font=scriptsize}
    \caption{Processing time over LoC.}\label{fig:pTime_vs_LoC}
\end{figure}

We expect LoC and processing time to exhibit some linear dependence. To verify this, we visualise the plot of processing time against LoC for all software projects in Figure~\ref{fig:pTime_vs_LoC}.
The figure shows a consistent pattern between processing time and LoC across all the software projects. The scatter of the points in the plot follows a pattern, where the LoC increase, the processing time also increases. This relationship is further demonstrated through the regression lines added in the figure. We can see that Sportspress has a near-perfect linear trend throughout all the data points. However, several outliers can also be seen in the figure, especially the one that belongs to Symphony. This outlier has a value of 2326 LoC and 50.09 seconds of processing time. This specific data point is far from the projected trends of all the software projects. We inspected the file representing that data point, which is FrameworkExtension.php. This file contains a lot of nested if-else conditions in most functions, which explains the longer processing time. The creation of CFG for nested if-else conditions takes a longer compared to simpler source code.

\subsubsection{Discussion}
Overall, this performance analysis shows that \DT\ is an efficient model which can scale without problems to larger code bases. In the worst case scenario it takes less than half an hour to analyse a 27MB project. Considering that this kind of vulnerability detection is an offline task that is performed only periodically on a whole project, this cost is negligible. On the other hand, the average processing time per file, under one second, is also sufficiently small to make it possible to run the model on individual files at each commit as part of a continuous integration pipeline.

\subsection{Classification Performance}

We established that \DT\ is usable in terms of execution performance. We now consider the usability in terms of vulnerability detection. As discussed above, we make the assumption that the popular projects currently do not have any security vulnerabilities. Hence we regard any positive reported by \DT\ as a false positive. In Table~\ref{tab:execution_performance} we report the accuracy of File-A and File-G for all projects.
As observed in Section~\ref{sec:classification}, File-G has fewer FP than File-A, which translates to a higher accuracy on a vulnerability-free dataset. Hence we recommend to use File-G especially if the code base is large and the priority is to reduce false positives.
We note a drop in accuracy for both models compared to the results reported in Table~\ref{tab:exp_filelevel} for \gitfi.
This is due to the model being trained on a different code base, therefore encountering novel unfamiliar coding patterns. 
Still \DT\ File-G has an average accuracy of 71.37\%, which we consider an encouraging result in terms of generalisation to a new dataset, especially in comparison with the poor generalisation ability of the other tools considered in Sections~\ref{sec:tools}.

\subsection{Vulnerability Detection}

Finally we attempt to use \DT\ to discover new vulnerabilities. 
We assume that the smaller plugins we considered may indeed contain vulnerabilities, as discussed above.
Our priority is to minimise the manual effort spent reviewing reported positives. Machine learning techniques make no promise of completeness, so it is preferable to miss some detections but focus the code reviewing efforts on code more likely to contain security flaws.

We follow a layered approach: we first use File-G (our practical model with fewer false positives) to detect potentially vulnerable files from the smaller plugins projects. That yields 177 potentially vulnerable files across two vulnerabilities, SQLi and XSS. 
Then, to better localise vulnerabilities, we apply Func-A (our most precise function-level model) on the 439 functions extracted from these 177 files. That yields 60 potentially vulnerable functions that are distributed across SQLi, XSS and OSCI vulnerabilities. Table~\ref{tab:novel_vulns} shows the results for the layered approach.

In absence of ground truth, we need to resort to manual inspection to verify the results. Several appeared suspicious (say concatenate a SQL string to a variable) but we did not have sufficient familiarity with the application to determine unequivocally if they constituted a vulnerability. We were able to confirm 4 of these functions as actual security vulnerabilities, and we responsibly disclosed our findings to the respective developers. We publicly disclosed 2 of them after they were patched and we describe them below.

\begin{table}[h]
    \scriptsize
%    \captionsetup{font=footnotesize}
    \caption{Layered approach results.}
    \label{tab:novel_vulns}
    \centering
    \begin{tabular}{p{2.5cm} p{0.4cm} p{1cm} p{0.9cm} p{1cm}}
    \toprule
    \textbf{Software project}& \textbf{TN} & \textbf{FP-SQLi} & \textbf{FP-XSS} & \textbf{FP-OSCI}\\
    \midrule
    \multicolumn{5}{c}{\textbf{A}: File-level granularity detection using File-G}\\
    \midrule
    Booking calendar & 9 & 7 & 0 & 0\\
    Payment form paypalpro & 7 & 1 & 5 & 0\\
    Paypal for digital goods & 3 & 4 & 0 & 0\\
    Sportspress & 123 & 31 & 102 & 0\\
    Simple Jobs Board & 171 & 4 & 23 & 0\\
    \midrule
    \multicolumn{5}{c}{\textbf{B}: Function-level granularity detection using Func-A}\\
    \midrule
    Booking calendar & 10 & 0 & 1 & 0\\
    Payment form paypalpro & 7 & 3 & 1 & 0\\
    Paypal for digital goods & 10 & 1 & 4 & 0\\
    Sportspress & 275 & 9 & 22 & 0\\
    Simple Jobs Board & 77 & 6 & 12 & 1\\
    \bottomrule
    \end{tabular}
\end{table}

\subsubsection{CVE-2020-14092} We found a SQL injection vulnerability in the plugin ``Payment Form for PayPal Pro''. It allowed any user to perform any SQL query they wanted, including retrieving user login information. This received a CVSS score of 9.8 (critical). Figure~\ref{fig:vul1} shows the vulnerable code snippet from the source codes. 
    
\begin{figure*}[!t]
    %\vspace{-2mm}
%    \captionsetup{font=footnotesize}
    \begin{center}
\hrule
\begin{lstlisting}[language=PHP,escapechar=@,basicstyle=\fontsize{8pt}{8pt}\selectfont\ttfamily]
function cp_ppp_init_ds(){
    @\hl{\$query\_result =  cp\_ppp\_ds( \$\_REQUEST );}@
    $err = mysqli_error( $cpcff_db_connect );
    if ( !is_null( mysqli_connect_error() )) 
       $err .= mysqli_connect_error();
    if ( $_REQUEST['cffaction'] == test_db_query){
       print_r( ( ( empty( $err ) ) ? $query_result:$err));
    } else {
       $result_obj = new stdClass;
       if( !empty( $err ) ){ 
          $result_obj->error = $err;
       } else { 
          $result_obj->data = $query_result
       }
       @\hl{print(json\_encode(\$result\_obj));}@
    }
} 
\end{lstlisting}
\hrule
    \end{center}
    \caption{SQLi vulnerability CVE-2020-14092.}\label{fig:vul1}
\end{figure*}
    
    \subsubsection{CVE-2020-13892} We found an XSS vulnerability in the ``SportsPress'' plugin, which allowed authenticated users to add malicious JavaScript to the WordPress installation. This received a CVSS score of 5.4 (medium). Figure~\ref{fig:vul2} shows the vulnerable code snippet from the source code.

\begin{figure*}[!t]
    %\vspace{-2mm}
%    \captionsetup{font=footnotesize}
    \begin{center}
\hrule
\begin{lstlisting}[language=PHP,escapechar=@,basicstyle=\fontsize{8pt}{8pt}\selectfont\ttfamily]
public function save(){
    parent::save();
    if ( isset( $_POST[ 'sportpress_events_teams_delimiter' ]))
       @\hl{update\_option( 'sportpress\_event\_teams\_delimiter', \$\_POST['sportpress\_event\_teams\_delimiter']);}@
}
\end{lstlisting}
\hrule
    \end{center}
    \caption{XSS vulnerability CVE-2020-13892.}\label{fig:vul2}
\end{figure*}

We tested the tools from Section~\ref{sec:tools} on these projects to see if they could detect either of the above vulnerabilities, but none succeeded.

\section{Conclusions}

We have presented \DT, a novel vulnerability detection approach which aims to capture contextual information from real-world vulnerabilities in order to reduce false positives and false negatives. Our approach combines a Gated Recurrent Unit to learn long term sequential dependencies of source code tokens and a Graph Convolutional Network to incorporate contextual information from the control flow graph. \DT\ exhibits scalable execution performance to tackle large source code bases, and achieves a better classification performance that the state-of-the-art on both synthetic and realistic datasets. Using \DT\ we were able to detect, with limited manual effort, 4 novel security vulnerabilities in WordPress plugins, which other detection tools failed to detect.

%\bibliographystyle{IEEEtran}
%\bibliography{references} 

% Inlined biblio for ArXiv (seriously guys?)


\end{document}